\documentclass[11pt,a4paper]{article}
\usepackage[margin=2.3cm]{geometry}
\usepackage[utf8]{inputenc}
\usepackage[T1]{fontenc}
\usepackage{lmodern}
\usepackage{amsmath,amssymb,bm}
\usepackage{graphicx}
\usepackage{booktabs}
\usepackage{xcolor}
\usepackage[numbers,sort&compress]{natbib}
\usepackage[colorlinks=true,linkcolor=blue!50!black,citecolor=blue!50!black,urlcolor=blue!50!black]{hyperref}
\usepackage{caption}
\newcommand{\bk}{\mathbf{k}}
\newcommand{\bq}{\mathbf{q}}
\newcommand{\bR}{\mathbf{R}}

\newcommand{\eps}{\varepsilon}
\newcommand{\Om}{\Omega}
\newcommand{\ee}{\mathrm{e}}
\newcommand{\ii}{\mathrm{i}}
\newcommand{\PAOFLOW}{\textsc{PAOFLOW}}
\newcommand{\EPW}{\textsc{EPW}}
\newcommand{\QE}{\textsc{Quantum ESPRESSO}}

\title{Electron--phonon interactions in a pseudo-atomic orbital basis: superconductivity, transport, spectroscopy and thermal emission from one interpolated vertex}
\author{Marco Buongiorno Nardelli$^{1,2,\ast}$\\[0.3em]
\small $^1$Department of Physics, University of North Texas, Denton, TX 76203, USA\\
\small $^2$Santa Fe Institute, Santa Fe, NM 87501, USA\\
\small $^\ast$mbn@unt.edu}
\date{}

\begin{document}
\maketitle

\begin{abstract}
Electron–phonon interactions govern superconductivity, electrical resistivity, optical absorption and electronic excitation lifetimes in solids. First-principles calculations require interpolating this coupling onto dense Brillouin-zone grids, traditionally using maximally localized Wannier functions that demand material-specific tuning. Here we show that a fixed basis of pseudo-atomic orbitals—automatically generated from pseudopotentials and implemented in the open-source code PAOFLOW—captures the electron–phonon vertex with equivalent accuracy and no manual tuning. Computed from density-functional perturbation theory or displaced supercells, and incorporating long-range fields and spin–orbit coupling, the vertex matches established codes on identical inputs. We validate the method on lead—reproducing its phonon dispersion, superconducting gap, critical temperature, resistivity, photoemission kink and cyclotron masses—and on silicon, capturing phonon-assisted optical absorption, wafer emissivity and carrier mobilities. This parameter-free framework enables high-throughput, automated electron–phonon calculations across large materials databases.
\end{abstract}

\section*{Introduction}

In the framework of density functional theory (DFT)~\cite{Martin2020}, the electron--phonon vertex, the matrix element between two electronic
states of the change of the self-consistent potential under a lattice
displacement, is the common input of a wide range of observables. The
superconducting gap and $T_c$, the phonon-limited resistivity and
carrier mobility, the infrared optical constants of a metal, the
phonon-assisted absorption of a semiconductor and the kinks that
photoemission resolves all derive from it. In practice,
density-functional perturbation theory (DFPT) provides the vertex on
coarse grids, at the cost of one linear-response calculation per
phonon wave vector\cite{Baroni2001}. On the other hand, each of
these observables is an integral over the Fermi surface or over the
band edges that converges only with $10^5$--$10^6$ points in both the
electron wave vector $\bk$ and phonon wave vector $\bq$. Bridging the two grids is the
central problem of first-principles electron--phonon calculations.

Since the work of Giustino, Cohen and
Louie~\cite{Giustino2007,Giustino2017}, this interpolation has been
carried out with maximally localized Wannier functions.
\EPW{}~\cite{Ponce2016,Lee2023}, \textsc{Perturbo}~\cite{Zhou2021},
\textsc{Phoebe}~\cite{Cepellotti2022},
\textsc{elphbolt}~\cite{Protik2022} and
\textsc{EPIq}~\cite{Marini2024} all rest on it, while \textsc{Abinit}
interpolates the perturbing potentials instead~\cite{Brunin2020}.
Wannier functions are exact for isolated groups of bands. However,
metals, heavy elements with spin--orbit coupling and entangled
conduction bands require a disentanglement window, a choice of
projections and a localization procedure whose outcome, and gauge,
change from material to material. As a consequence, the same
calculation cannot be set up automatically across a database of
compounds.

Since the introduction of the projection method by Buongiorno Nardelli and co-workers in 2013~\cite{Agapito2013,2016-LuisAgapito-PRB-93-35104,2016-LuisAgapito-PRB-93-125137}, later implemented in the \PAOFLOW{} code~\cite{BuongiornoNardelli2018,Cerasoli2021,Jayaraj2026}, pseudo-atomic orbitals (PAOs) have become a natural alternative to Wannier functions: by constructing tight-binding Hamiltonians from the projections of the Kohn--Sham wavefunctions on such states, the basis is
fixed by the element, the projectability of each state measures how
well the basis represents it, and no localization or disentanglement
is involved (Methods). Although Agapito and Bernardi proposed to apply this framework to the interpolation of the electron--phonon vertex~\cite{Agapito2018}, such initial proposal did not result in a new systematic approach.

In this paper we discuss the PAO construction of the vertex and use
it to compute, from one interpolated vertex, a broad set of
observables, including some that have received little attention from
first principles. In particular, ab initio calculations of the
thermal emissivity of a semiconductor wafer and of a metal as a
function of temperature are scarce, and so are those of the
photoemission spectral function of a metal and its kink.

The DFPT vertex is built natively from the
DFPT potentials and the wavefunctions, interpolated in
both the electron and the phonon wave vector, with the proper treatment of the 
long-range dipole and quadrupole fields. A second
source of the same vertex is the finite differences of the projected
Hamiltonian of displaced supercells in the fixed orbital basis.
The latter does not require a linear-response code and is defined at any reference
geometry. Its closest relatives take the difference of the
potential~\cite{Dai2026} or of the spectral representation of the
displaced Hamiltonian~\cite{Poliukhin2026} between the unperturbed
Bloch states, and then interpolate with Wannier functions. Finally,
the coarse coupling written by \EPW{} can be read in place of the
DFPT potentials, so that the two interpolations can be
compared on identical inputs (Supplemental Information).

Lead, the textbook
strong-coupling superconductor, and silicon, the most well-characterized of semiconductors, are the prototype systems throughout this work. In lead, the Eliashberg function is known
from tunnelling~\cite{McMillanRowell1965}, phonons and
resistivity have been measured with high precision and the
electron--phonon kink has been observed directly in
photoemission~\cite{Reinert2003}. It is also a case where its
scalar-relativistic (SR) description fails by a factor of
two~\cite{Heid2010,Goudreault2025} and can provide a benchmark on the SR vs. spin--orbit coupling (SOC) routes. As a first step we have validated
the vertex against \EPW{} and the finite-difference route against
DFPT. We then show that a single set of converged calculations with
spin--orbit coupling reproduces the phonons, the gap, the resistivity,
the infrared optics, the photoemission kink and the cyclotron masses
of lead. Finally, we turn to the absorption, thermal emission and
carrier mobilities of silicon, directly compared with
experiments.

\section*{Results}

\subsection*{The electron--phonon vertex in the pseudo-atomic orbital basis}
\label{sec:vertex}

\begin{figure}[t]
\centering
\includegraphics[width=0.78\linewidth]{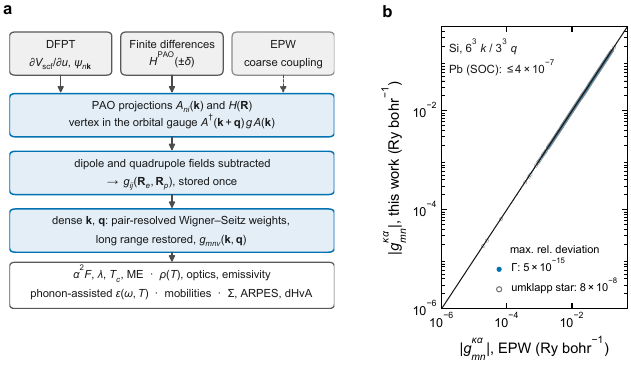}
\caption{\textbf{The construction and its validation.} \textbf{a}, Path from the sources of the vertex (DFPT potentials, finite differences of the projected Hamiltonian of displaced supercells, coarse coupling of \EPW{}) through the PAO gauge and the real-space vertex $g(\bR_e,\bR_p)$ to the dense-grid observables. \textbf{b}, Native coarse vertex of Si ($6^3$ $k$, $3^3$ $q$, 12 bands) compared with that of \EPW{} on identical inputs, for every band pair, mode and $\bk$ point at $\Gamma$ (filled) and at a star member reached through an umklapp vector (open).}
\label{fig:method}
\end{figure}

The construction starts from a DFT calculation on a coarse $\bk$
grid. The Kohn--Sham states $\psi_{n\bk}$ are
projected on $N_{\rm awf}$ pseudo-atomic orbitals $\phi_i$ defined as
radial functions times spherical harmonics generated from the
pseudopotential of each element, with projections
$A_{ni}(\bk) = \langle\psi_{n\bk}|\phi_{i\bk}\rangle$, and the bands
whose projectability $p_{n\bk} = \sum_i|A_{ni}(\bk)|^2$ exceeds a
threshold define the PAO Hamiltonian~\cite{Agapito2013} (Methods). Its Fourier transform is the
real-space Hamiltonian, which is interpolated to any $\bk$ with
pair-resolved Wigner--Seitz weights. The DFPT vertex
$g^{\kappa\alpha}_{mn}(\bk,\bq) =
\langle\psi_{m,\bk+\bq}|\partial_{\bq\kappa\alpha}V|\psi_{n\bk}\rangle$,
the matrix element between Kohn--Sham states of the derivative of the
self-consistent potential with respect to the displacement of atom
$\kappa$ along the Cartesian direction $\alpha$, is rotated into the
same orbital basis with the same projections,
\begin{equation}
g^{{\rm PAO},\kappa\alpha}_{ij}(\bk,\bq) = \sum_{mn} A^*_{mi}(\bk+\bq)\, g^{\kappa\alpha}_{mn}(\bk,\bq)\, A_{nj}(\bk),
\label{eq:gpao}
\end{equation}
where $i$ and $j$ label the orbitals (Methods). This rotation is exact
within the span of the projected bands and independent of the band
phases, since every band index appears once as a bra and
once as a ket. The double Fourier transform of Eq.~\eqref{eq:gpao} to
$(\bR_e,\bR_p)$ is the object that is stored and interpolated. The
band- and mode-resolved coupling on the dense grids then follows by
rotating back with the eigenvectors of the interpolated Hamiltonian
and with the phonon eigenvectors of the interpolated dynamical matrix
(Methods), as summarized in Fig.~\ref{fig:method}a.

Three ingredients make the construction general. First, the vertex is
built natively from the DFPT potentials, as the sum of the
induced Hartree and exchange--correlation response, the bare local
term and the Kleinman--Bylander derivative terms, in the gauge of the
projections, so that no \EPW{} run is required. The irreducible $\bq$
points of DFPT are unfolded by rotating the scalar potentials on the
FFT grid, and a DFT calculation on the irreducible wedge is unfolded
by rotating the wavefunctions, with SU(2) and time reversal for
spinors. Second, the long-range dipole (Fr\"ohlich) and quadrupole
fields, which are not smooth in $\bq$, are subtracted on the coarse
grid as the Ewald sums of Refs.~\cite{Verdi2015,Brunin2020} and
restored at every dense $\bq$. In particular, in the PAO gauge the
long-range vertex is independent of $\bk$ and diagonal in the
orbitals. Finally, the Wigner--Seitz images are assigned per orbital
pair, at the image of $\bR + \bm\tau_j - \bm\tau_i$ closest to the
origin. Assigning the image by the bond vector, rather than by the lattice vector {\bf R} alone, preserves the symmetry of the interpolated Hamiltonian and vertex in cells with more than one atom.

In order to validate the construction, we have compared the coarse
vertex with that of \EPW{} on identical inputs. In
Fig.~\ref{fig:method}b we show the two vertices of Si for every band
pair, mode and $\bk$ point. The largest relative deviation is
$5\times10^{-15}$ at $\Gamma$ and $8\times10^{-8}$ at a star member
reached through an umklapp vector. Over all the coarse $q$ points the
deviation is of the order of $10^{-7}$ for Si, $2\times10^{-6}$ for
scalar-relativistic Pb and $4\times10^{-7}$ for Pb with spin--orbit
coupling, apart from $3$--$5\times10^{-4}$ at the umklapp $\bk$ points
of the two largest stars of Si, a residual of \EPW{} itself
(Supplementary Information). With the same $12^3/6^3$ coarse coupling
and the same dense grids, the interpolated phonon-assisted absorption
of Si is 1.03 times that of \EPW{} on average (1.00--1.09 over the
spectrum; Supplementary Information).  These comparisons give us
confidence in the interpolated vertex.

\subsection*{A second source of the vertex: finite differences in a fixed basis}
\label{sec:fd}

\begin{figure}[t]
\centering
\includegraphics[width=\linewidth]{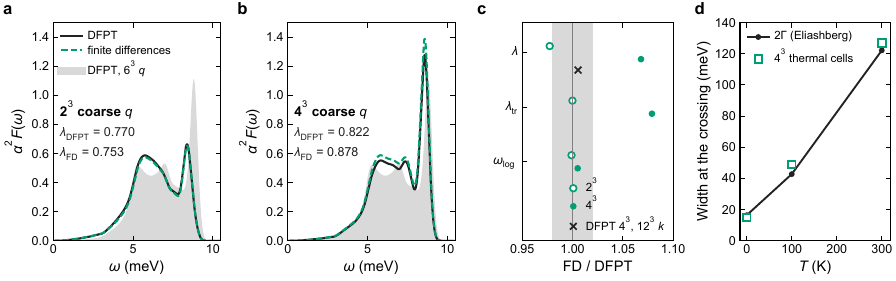}
\caption{\textbf{Electron--phonon vertex from finite differences in the fixed orbital basis} (lead, scalar relativistic). \textbf{a},\textbf{b}, Eliashberg functions from the finite-difference (dashed) and DFPT (solid) vertices on the same $2^3$ (\textbf{a}) and $4^3$ (\textbf{b}) coarse $q$ grids, both interpolated to the $60^3/60^3$ dense grids; the shaded curve is the $6^3$ DFPT production. \textbf{c}, Ratios of $\lambda$, $\lambda_{\rm tr}$ and $\omega_{\log}$ from the two vertices at equal sampling ($2^3$, open; $4^3$, filled); the cross refers to DFPT on the $4^3$ grid with the $12^3$ $k$ grid of the supercells instead of $16^3$, and the gray band to $\pm2$\%. \textbf{d}, Width of the Fermi-crossing state in the unfolded spectral function of $4^3$ special-displacement cells (three configurations per temperature) compared with the static Eliashberg linewidth $2\Gamma$ on the same $q$ grid.}
\label{fig:fd}
\end{figure}

The DFPT potentials are one way to obtain the vertex, and the
projected Hamiltonian provides another, which doesn't requires a
linear-response code. Since the orbitals are kept at the reference
positions, the matrix of the Kohn--Sham Hamiltonian in that basis can
be differentiated term by term. The central difference of the
projected Hamiltonian of a supercell with one atom displaced by
$\pm\delta$ is then the matrix of $\partial V/\partial u$ in the
reference orbitals, i.e., the object of Eq.~\eqref{eq:gpao} restricted
to the projected bands (Methods). For lead, the derivative with
respect to the reference atom is obtained from six DFT calculations of
a $2\times2\times2$ cell (eight atoms, $\delta = 0.06$~bohr). The
derivatives with respect to the other atoms then follow from
translation symmetry, and the derivative folds to the primitive cell
with $\bq$ on the $2^3$ grid of the supercell. From this point on, the
real-space vertex $g(\bR_e,\bR_p)$ and the dense-grid drivers are the
same as in the DFPT route.

In Fig.~\ref{fig:fd} we compare the two sources of the vertex at equal
sampling. Interpolated to the $60^3/60^3$ dense grids from the same
$2^3$ coarse $q$ grid, we obtain $\lambda = 0.753$ and 0.770 from the
finite-difference and DFPT vertices, respectively
(Fig.~\ref{fig:fd}a,c), the same $\omega_{\log}$ and
$\lambda_{\rm tr}$, and transport functions that coincide within 2\%
point by point. On the matrix elements between states within 0.5~eV of $E_F$, the finite-difference vertex and the DFPT vertex restricted to the same projected bands agree to 0.1\% (Supplementary Information). With a $4\times4\times4$
cell (64 atoms) we obtain the vertex on the $4^3$ $q$ grid, with
$\lambda = 0.878$, $\omega_{\log} = 6.12$~meV and
$\lambda_{\rm tr} = 0.710$, compared with 0.822, 6.09~meV and 0.658
for DFPT on the same grid (Fig.~\ref{fig:fd}b,c). At equal sampling
the two sources thus agree on $\omega_{\log}$ to 0.5\% and on
$\lambda$ within $-2$ and $+7$\%, which sets the present accuracy of
the route. We have verified that the difference at $4^3$ is due
neither to the convergence threshold of the supercell calculations,
which leaves $\lambda$ of the $2\times2\times2$ cells unchanged, nor
to their $\bk$ sampling, which changes the DFPT $\lambda$ on the $4^3$
grid by 0.5\% (Fig.~\ref{fig:fd}c and Supplementary Information).
Random all-atom displacements with a local least-squares extraction
replace the Cartesian set at the same accuracy (force constants within
2.1\% and $\lambda$ within 0.1\% on the ideal cell), which makes the
route affordable for large cells.

It is important to note that the finite difference construction
uses only eigenvalues and eigenvectors, so that it
applies to any DFT Hamiltonian, including
hybrid functionals and DFT+$U(+V)$. Moreover, it is defined for any
reference geometry, such as a thermal cell built with the special
displacement method~\cite{Zacharias2016,Zacharias2020}, where the
vertex contains the multiphonon terms as a thermal average. As a test,
we have computed the width of the Fermi-crossing state in the unfolded
spectral function of $4^3$ special-displacement cells of lead. The
widths, 15, 49 and 127~meV at 0, 100 and 300~K, follow the static
Eliashberg linewidth $2\Gamma$ on the same $q$ grid (16, 43 and
122~meV), as shown in Fig.~\ref{fig:fd}d. The static disorder of the
thermal cells thus reproduces the Williams--Lax limit of the
self-energy on the same $q$ grid.

\subsection*{Lead: phonons and Eliashberg function}
\label{sec:pb_sc}

\begin{figure}[t]
\centering
\includegraphics[width=\linewidth]{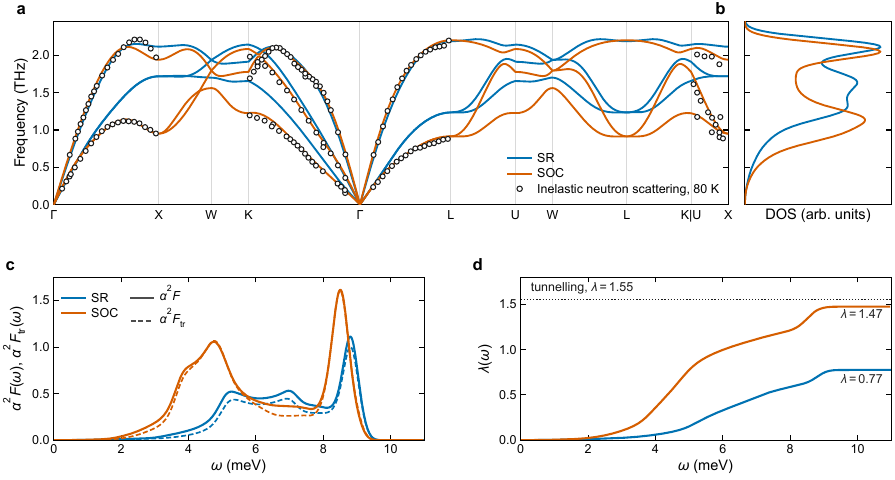}
\caption{\textbf{Phonons and Eliashberg functions of lead}, scalar relativistic (SR, blue) and with spin--orbit coupling (SOC, orange). \textbf{a}, Phonon dispersion from DFPT on a $6^3$ $q$ grid compared with the neutron frequencies at 80~K of Stedman \textit{et al.}~\cite{Stedman1967} (circles). \textbf{b}, Phonon density of states. \textbf{c}, Isotropic (solid) and transport (dashed) Eliashberg functions on the $60^3/60^3$ dense grids. \textbf{d}, Cumulative coupling $\lambda(\omega) = 2\int_0^\omega \alpha^2F(\omega')/\omega'\,d\omega'$; the dotted line indicates the tunnelling value~\cite{McMillanRowell1965}.}
\label{fig:pb_a2f}
\end{figure}

The calculated properties of lead are collected in Table~\ref{tab:pb}.
In Fig.~\ref{fig:pb_a2f}a we show the phonon dispersion from DFPT on a
$6^3$ $q$ grid, scalar relativistic and with spin--orbit coupling, together with the neutron frequencies of Stedman \textit{et
al.}~\cite{Stedman1967}. With spin--orbit coupling the calculated
branches fall on the measured ones along all three directions. In
particular, the [100] transverse branch rises to 1.14~THz at
$q = 0.68$ (1.12~THz at 0.65 measured) and dips to 0.95~THz at X, as
measured, the [111] transverse branch reaches 0.91~THz at L (0.90~THz
measured), and the three branches at K lie at 1.23, 1.78 and 2.09~THz,
while the measured ones are at 1.20, 1.69 and 1.98~THz.

On the other hand, in the scalar-relativistic calculation the longitudinal branches are within 8\% of experiment,
but the transverse zone-boundary modes are too hard by 37\% at L, 38\%
at K and 81\% at X (1.72 instead of 0.95~THz). This is the known
failure that spin--orbit coupling repairs~\cite{Heid2010}.

The Eliashberg function follows the phonons, as shown in
Fig.~\ref{fig:pb_a2f}c. With spin--orbit coupling the transverse peak
is located at 4.7~meV with a shoulder at 3.8~meV, where the tunnelling
$\alpha^2F$ has its first peak (4.4~meV), and the longitudinal peak at
8.5~meV coincides with the tunnelling one. As a result, 68\% of
$\lambda$ comes from the modes below 6~meV, while only 43\% does in
the scalar-relativistic function, whose first structures lie at 5.3
and 7.0~meV (Fig.~\ref{fig:pb_a2f}d). The coupling constant
$\lambda = 1.47$ is converged to 0.5\% between the $72^3/36^3$ and
$60^3/60^3$ dense grids and varies by $\pm2$\% between Fermi-surface
smearings of 0.03 and 0.08~eV. This value is 5\% below the tunnelling
inversion (1.55)~\cite{McMillanRowell1965} and close to earlier DFPT
calculations with spin--orbit coupling~\cite{Heid2010}, while
$\omega_{\log} = 5.06$~meV (59~K) compares well with 4.8~meV
(56~K)~\cite{AllenDynes1975}. In the scalar-relativistic case we find
instead $\lambda = 0.77$ and $\omega_{\log} = 6.1$~meV.

In order to validate the construction on lead, we have compared the
native coarse vertex with the coarse coupling that \EPW{} computes from
the same DFPT run, which \PAOFLOW{} can also read as an
alternative source of the vertex (Methods). The two vertices agree to
$4\times10^{-7}$ with spin--orbit coupling and to $2\times10^{-6}$
without (Supplementary Information). This implies that the value of $\lambda$
is determined by the DFPT, namely the $\bk$ grid, the lattice constant
and the treatment of spin--orbit coupling, and by the convergence of
the dense grids, not by the interpolation. The latter is not a minor
issue: on our coupling the $48^3/24^3$ dense grids change $\lambda$ by
$-0.2$\% without spin--orbit coupling but by $+13$\% with it (1.66
instead of 1.47). This is due to the $24^3$ $q$ grid, too coarse to sample the strongly peaked coupling of the soft transverse modes with spin–orbit coupling: the $72^3/36^3$ grids agree with $60^3/60^3$ to 0.5\%. Converged grids
on both sides of the interpolation are therefore needed before the
effect of spin--orbit coupling can be isolated.

\begin{table}[t]
\caption{\textbf{Calculated properties of lead}, scalar relativistic (SR) and with spin--orbit coupling (SOC), on the $12^3/6^3$ coarse and $60^3/60^3$ dense grids, compared with experiment. The emissivity is the total hemispherical value of the opaque metal, with the interband term in parentheses.}
\label{tab:pb}
\small
\begin{tabular}{p{0.31\linewidth}p{0.16\linewidth}p{0.24\linewidth}p{0.16\linewidth}}
\toprule
quantity & SR & SOC & experiment \\
\midrule
$\lambda$ & 0.77 & 1.47 & 1.55 (tunnelling) \\
$\omega_{\log}$ (meV) & 6.1 & 5.06 & $\approx 4.8$ \\
$T_c$, Allen--Dynes, $\mu^* = 0.10$ (K) & 3.2 & 7.30 & 7.19 \\
$T_c$, isotropic ME, gap closure / linearized (K) & 3.8 / 3.7 & 7.87 / 7.80 & 7.19 \\
$T_c$, anisotropic ME, gap closure / linearized (K) & 4.1 / 4.1 & 8.14 / 8.13 & 7.19 \\
$\Delta_0$ (gap edge); anisotropic range (meV) & 0.60; 0.52--0.72 & 1.52; 1.36--1.64 & 1.33--1.40 \\
$2\Delta_0/k_BT_c$ & 3.6--3.8 & 4.5 & 4.3--4.5 \\
$\lambda_{\rm tr}$, $\hbar\omega_p$ (eV) & 0.62, 8.09 & 1.33, 8.95 & ---, 7.4 (Drude fit) \\
$\rho(300\,\mathrm{K})$: $\alpha^2F_{\rm tr}$; SERTA / MRTA ($\mu\Omega$\,cm) & 11.4; 13.6 / 11.2 & 20.0; 21.0 / 19.6 & 21.3 \\
emissivity, 300~K & 0.027 (0.027) & 0.044 (0.046) & 0.057--0.075 \\
phonons at X, T / L (THz) & 1.72 / 2.12 & 0.95 / 1.94 & 0.95 / 1.97 \\
ARPES: slope $\lambda$ at $k_F$; $\mathrm{Re}\,\Sigma$ maximum & 0.67; 6.1 meV at 7.5 meV & 1.59; 8.5 meV at 6.8 meV & 1.55; $\approx 7$ meV \\
dHvA frequencies $\psi_1$ / $\zeta$ / $\nu$ / $\eta$ at $[001]$ (kT) & 14.8 / 1.65 / --- / --- & 19.7 / 2.06 / 4.94 / 23.0 & 22.2 / 2.54 / 5.38 / 24.4 \\
cyclotron masses $m^*_{\rm band}(1+\lambda_{\rm orbit})$: $\psi_1$, $\zeta$, $\nu$, $\eta$ at $[001]$ ($m_e$) & --- & 1.63, 0.68, 1.27, 2.66 (band: 0.71, 0.26, 0.50, 1.05) & 1.58--1.64, 0.68--0.75, 1.20--1.22, 2.59 \\
\bottomrule
\end{tabular}
\end{table}

\subsection*{Lead: the superconducting gap}
\label{sec:pb_me}

\begin{figure}[t]
\centering
\includegraphics[width=\linewidth]{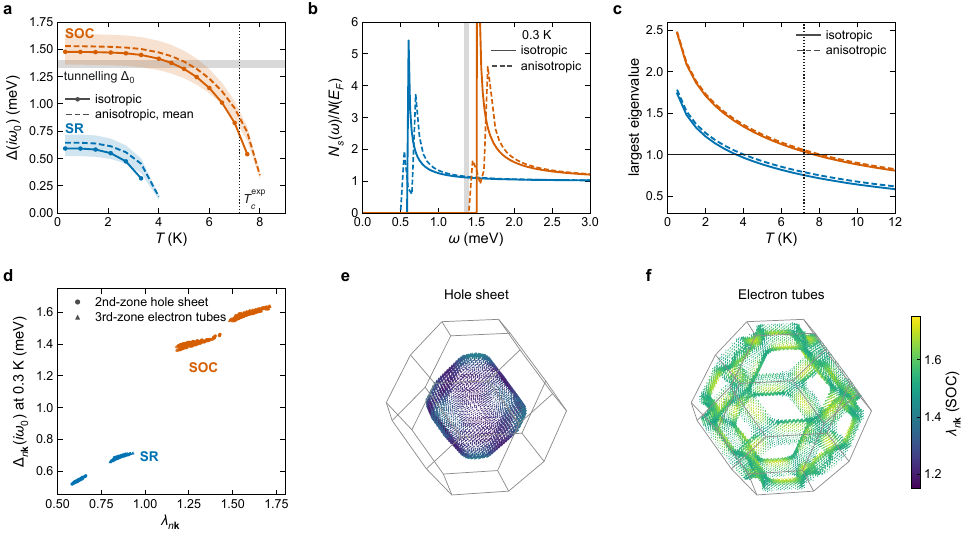}
\caption{\textbf{Superconducting gap of lead} from the Migdal--Eliashberg equations ($\mu^* = 0.10$, $60^3/60^3$ dense grids), SR (blue) and SOC (orange). \textbf{a}, Gap $\Delta(\ii\omega_0)$ as a function of temperature from the isotropic equations (circles) and mean (dashed) and range (shaded) of $\Delta_{n\bk}(\ii\omega_0)$ from the anisotropic equations; the gray band and the dotted line indicate the measured $\Delta_0$ and $T_c$. \textbf{b}, Quasiparticle density of states at 0.3~K. \textbf{c}, Largest eigenvalue of the linearized kernel; $T_c$ is the temperature at which it crosses one. \textbf{d}, Gap $\Delta_{n\bk}(\ii\omega_0)$ at 0.3~K as a function of $\lambda_{n\bk}$ for every irreducible Fermi-surface state. \textbf{e},\textbf{f}, $\lambda_{n\bk}$ on the second-zone hole sheet and on the third-zone electron tubes of the SOC Fermi surface, in the first Brillouin zone.}
\label{fig:pb_me}
\end{figure}

To determine the gap and $T_c$, we have solved the isotropic and the
anisotropic Migdal--Eliashberg equations on the Matsubara axis and
continued them to the real axis with Pad\'e approximants and with the
analytic-continuation equations (Methods). The results are summarized
in Fig.~\ref{fig:pb_me}. The screened Coulomb repulsion between the
electrons of a pair enters the equations through the Coulomb
pseudopotential $\mu^*$. With $\mu^* = 0.10$ we obtain from the
isotropic equations a gap edge of 1.52~meV at 0.3~K and $T_c = 7.80$~K
from the linearized kernel (7.87~K from the closure of the gap), while
the measured values are 1.33--1.40~meV and 7.19~K, respectively. The gap ratio
$2\Delta_0/k_BT_c = 4.5$ reproduces the strong-coupling value of lead,
4.3--4.5. In the anisotropic case the equations act on the pair
coupling of the 1046 irreducible Fermi-surface states of the $60^3$
grid. In Fig.~\ref{fig:pb_me}e,f we show $\lambda_{n\bk}$ on the two
sheets of the Fermi surface: it ranges from 1.18 to 1.42 on the
second-zone hole sheet and from 1.48 to 1.71 on the third-zone
electron tubes, and the gap follows it state by state
(Fig.~\ref{fig:pb_me}d). The gap spreads by $-11$ and $+7$\% about its
mean, and $T_c = 8.13$--$8.14$~K. Lead is thus a weakly anisotropic
single-gap superconductor, as expected, and the isotropic and
anisotropic routes differ by about 0.3~K.

The residual excess of 0.6--1.0~K over experiment lies within the
uncertainty of $\mu^*$. Indeed, on the same $\alpha^2F$ and pair
coupling, for $\mu^* = 0.10$, 0.13 and 0.15 we find $T_c = 7.80$, 7.32
and 7.03~K from the isotropic linearized kernel (7.87, 7.39 and 7.06~K
from the closure of the gap) and 8.13, 7.76 and 7.54~K from the
anisotropic one (8.14, 7.78 and 7.62~K), with gap edges of 1.52, 1.41
and 1.35~meV and a gap ratio that stays at 4.4--4.5 (Supplementary
Table~S1). The tunnelling value
$\mu^* = 0.131$~\cite{McMillanRowell1965} therefore brings the
isotropic route to 7.3~K and the anisotropic one to 7.8~K, 0.1--0.6~K
above experiment, while the gap edge moves to 1.41~meV, close to the
measured 1.33--1.40~meV. On the other hand, without spin--orbit
coupling the gap is 0.60~meV and $T_c = 3.7$--$4.1$~K.

\subsection*{Lead: resistivity, infrared optics and thermal emission}
\label{sec:pb_tr}

\begin{figure}[t]
\centering
\includegraphics[width=\linewidth]{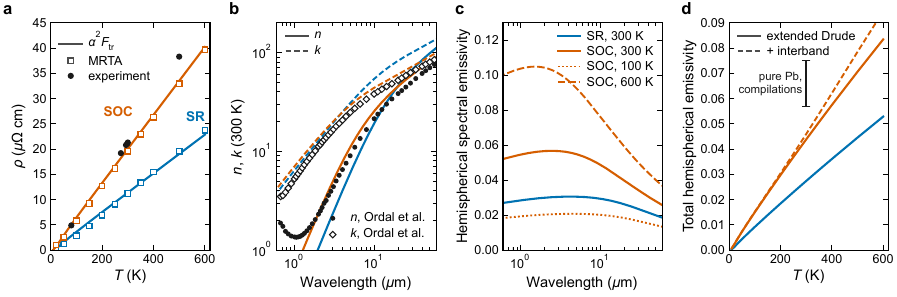}
\caption{\textbf{Transport, infrared optics and thermal emission of lead}, SR (blue) and SOC (orange). \textbf{a}, Phonon-limited resistivity from $\alpha^2F_{\rm tr}$ (lines) and from the state-resolved MRTA rates on the $60^3/60^3$ grids (squares) compared with the measured values of the CRC Handbook~\cite{Lide2003} (circles). \textbf{b}, Refractive index $n$ (solid) and extinction coefficient $k$ (dashed) at 300~K from the extended Drude model compared with Ordal \textit{et al.}~\cite{Ordal1987}. \textbf{c}, Hemispherical spectral emissivity of the opaque metal, SOC at 100, 300 and 600~K and SR at 300~K. \textbf{d}, Total hemispherical emissivity as a function of temperature from the extended Drude model (solid) and, for SOC, with the interband dielectric function added (dashed); the bar indicates the range of the compilations for pure unoxidized lead near room temperature.}
\label{fig:pb_tr}
\end{figure}

In Fig.~\ref{fig:pb_tr} we show the transport and optical properties
that follow from the same vertex. From the transport Eliashberg
function we obtain $\lambda_{\rm tr} = 1.33$ with spin--orbit coupling
and 0.62 without it. The ratio $\lambda_{\rm tr}/\lambda = 0.91$ (0.80
in the scalar-relativistic case) reflects the large-angle scattering
by the soft zone-boundary modes. The resistivity from Allen's
formula~\cite{Allen1971} follows the measured resistivity tabulated in
the CRC Handbook of Chemistry and Physics~\cite{Lide2003} within 6\%
up to 300~K (20.0 \textit{vs.} 21.3~$\mu\Omega$\,cm) and within 4\% at
80~K, and
falls 13\% short at 500~K. This is due to thermal expansion and
anharmonicity, which bend the measured curve upward and which a
harmonic calculation at a fixed lattice cannot
describe~\cite{Goudreault2025}. The state-resolved route, i.e., the
scattering rates of the Fermi-surface states in the self-energy and
momentum relaxation-time approximations (SERTA and MRTA) and the
relaxation-time conductivity, leads to 21.0 and 19.6~$\mu\Omega$\,cm
at 300~K, with the $48^3/48^3$ and $60^3/60^3$ grids within 1.5\% of
each other. In the scalar-relativistic case the three routes predict
11.2--13.6~$\mu\Omega$\,cm, a factor of two, as also found in
Ref.~\cite{Goudreault2025}.

The infrared optical constants of Fig.~\ref{fig:pb_tr}b are obtained
from the frequency-dependent conductivity of the extended Drude model,
with Allen's optical scattering rate and mass enhancement built from
$\alpha^2F_{\rm tr}$~\cite{Allen1971,Holstein1964} and with the
Fermi-surface plasma frequency (8.95~eV, while the Drude fit of Ordal
\textit{et al.} yields 7.4~eV). The refractive index is $n = 2.4$,
11.0, 25.5 and 45.6 at 2, 5, 10 and 20~$\mu$m, compared with 2.3, 9.0,
21.4 and 38.4 measured~\cite{Ordal1987}, and the extinction
coefficient $k$ is 15--24\% higher than measured. The normal-incidence
reflectivity is 0.953--0.969 between 2 and 20~$\mu$m. Finally, the
spectral emissivity at any angle (Fig.~\ref{fig:pb_tr}c) and the
Planck-weighted total hemispherical emissivity of the opaque metal
(Fig.~\ref{fig:pb_tr}d) are obtained via Kirchhoff's law. At 300~K we
find 0.044, and 0.046 when the interband dielectric function of the
extended basis is added, while the compilations for pure unoxidized
lead report 0.057--0.075, a spread due to surface finish and oxidation
that is of the same size as the remaining difference. As expected, the
emissivity rises with the resistivity as $\sqrt{\rho}$, from 0.011 at
75~K to 0.071 at 500~K. The scalar-relativistic value at 300~K, 0.027,
is again low by the factor of the coupling.

\subsection*{Lead: photoemission and quantum oscillations}
\label{sec:pb_spec}

\begin{figure}[t]
\centering
\includegraphics[width=\linewidth]{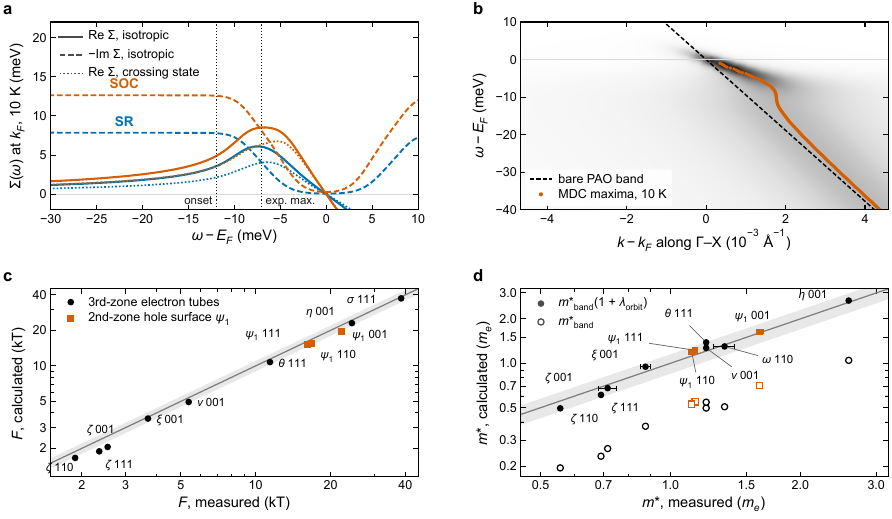}
\caption{\textbf{Electron--phonon renormalization of lead seen by photoemission and by quantum oscillations.} \textbf{a}, Self-energy at the Fermi crossing of $\Gamma$--X at 10~K, SR (blue) and SOC (orange): isotropic $\mathrm{Re}\,\Sigma$ (solid) and $-\mathrm{Im}\,\Sigma$ (dashed) from $\alpha^2F$, and $\mathrm{Re}\,\Sigma$ of the anisotropic crossing state (dotted); the vertical lines indicate the measured maximum of $\mathrm{Re}\,\Sigma$ and the onset of the renormalization~\cite{Reinert2003}. \textbf{b}, Photoemission map (SOC, He~I$\alpha$ 21.2~eV, 2~meV resolution) of the segment across $k_F$ dressed with the isotropic $\Sigma$ at 10~K, with the bare PAO band (dashed) and the maxima of the momentum distribution curves (MDCs; dots). \textbf{c}, de Haas--van Alphen frequencies of the symmetry orbits of the SOC Fermi surface for fields along [001], [111] and [110] compared with the measurements of Anderson and Gold~\cite{AndersonGold1965}. \textbf{d}, Cyclotron masses $m^*_{\rm band}(1 + \lambda_{\rm orbit})$ (filled) and bare band masses (open) compared with the Azbel'--Kaner cyclotron masses compiled by Anderson and Gold~\cite{AndersonGold1965} (Khaikin and Mina~\cite{Khaikin1962,Mina1964}, Young, Phillips). Squares refer to the second-zone hole surface $\psi_1$ and circles to the third-zone electron tubes; the gray bands indicate $\pm10$\% (\textbf{c}) and $\pm15$\% (\textbf{d}).}
\label{fig:pb_spec}
\end{figure}

Photoemission resolves the electron--phonon interaction directly, as
the kink of the quasiparticle dispersion within the phonon energies of
$E_F$ and the step of the linewidth beyond them. In order to compare
with experiment, we have simulated the band map of a segment of
$\Gamma$--X across the Fermi crossing with the photoemission matrix
elements of the PAO orbitals and a plane-wave final
state~\cite{Day2019} (Methods). The bands are dressed with the
Engelsberg--Schrieffer self-energy of
$\alpha^2F$~\cite{Engelsberg1963}, with the self-energy of the
anisotropic Fermi-surface state nearest the crossing, and with the
band-resolved Fan--Migdal self-energy on a dense $q$ grid.

In Fig.~\ref{fig:pb_spec}a,b we show the self-energy at the Fermi
crossing and the dressed band map with spin--orbit coupling. From the
slope of $\mathrm{Re}\,\Sigma$ at $E_F$ we obtain $\lambda = 1.59$ for
the crossing state and 1.65 for the isotropic form, in good agreement
with the value of 1.55 measured on the bulk band of
Pb(110)~\cite{Reinert2003}. The whole Fermi-surface range of the
slope, 1.31--1.92, brackets the measurement, whose $\bk$ point is not
the $\Gamma$--X crossing. $\mathrm{Re}\,\Sigma$ peaks at 8.5~meV for
$\omega = -7$~meV, in agreement with the measured maximum at
$\approx 7$~meV and with the onset of the renormalization below
$\approx 12$~meV, and $-\mathrm{Im}\,\Sigma$ rises by 12.4~meV between
$E_F$ and the top of the phonon spectrum, close to the measured
$\approx 13$~meV.

In the dressed map the maxima of the momentum distribution curves
(MDCs)
follow a band 2.2 times flatter than the bare one within 5~meV of
$E_F$, where the limit $1 + \lambda = 2.5$ is softened by the 2~meV
resolution, and recover the bare velocity beyond 15~meV. At 300~K the
linewidth reproduces $\pi\lambda k_BT$ to 0.6\%, the high-temperature
formula that Reinert \textit{et al.} have used to extract $\lambda$
from the temperature dependence. The Fan--Migdal form on an $18^3$ $q$
grid reproduces that linewidth within 15\%, but it does not resolve
the kink of a few meV, which requires the Fermi-surface forms. On the
contrary, without spin--orbit coupling the slope is 0.67--0.84 and the
maximum 6~meV, i.e., the kink is half the measured one. This result
confirms that spin--orbit coupling doubles the kink of lead, as it
doubles $\lambda$ and the resistivity.

The same anisotropic coupling determines the quantities that are measured in
quantum-oscillation experiments, i.e., the cyclotron
mass of each extremal orbit,
$m^* = m^*_{\rm band}(1 + \lambda_{\rm orbit})$, where
$\lambda_{\rm orbit}$ is the average of $\lambda_{n\bk}$ along the
orbit and the frequencies are those of the bare Fermi surface. All
thirteen symmetry orbits of the model of Anderson and
Gold~\cite{AndersonGold1965} are found on the PAO Fermi surface
(Fig.~\ref{fig:pb_spec}c and Supplementary Table~S2). Their
frequencies are 3--11\% below the measured ones for the large orbits
(19.7 \textit{vs.} 22.2~kT for the hole surface $\psi_1$ at $[001]$
and 23.0 \textit{vs.} 24.4~kT for $\eta$) and 12--20\% below for the
thin cross-sections $\zeta$ of the third-zone tubes. Without
spin--orbit coupling, instead, the hole surface and the tubes turn out
to be 33\% too small. The bare band masses are, in fact, 2.1--2.9 times smaller than the
measured cyclotron masses. 

In contrast,
$m^*_{\rm band}(1 + \lambda_{\rm orbit})$ reproduces the eleven
measured masses within $\pm15$\% with no systematic offset, as shown
in Fig.~\ref{fig:pb_spec}d. In particular, the calculated masses of
the $\psi_1$, $\zeta$, $\nu$ and $\eta$ orbits at $[001]$ are 1.63,
0.68, 1.27 and 2.66~$m_e$, and the measured ones 1.58--1.64,
0.68--0.75, 1.20--1.22 and 2.59~$m_e$, respectively. The
orbit-resolved enhancement is 1.24--1.31 on the hole surface and
1.53--1.60 on the tubes, the sheet averages of the $\lambda_{n\bk}$
distribution that sets the anisotropic gap (Fig.~\ref{fig:pb_me}e,f).
Anderson and Gold inferred the same two-sheet ratio of the mass
enhancement (1.95 and 2.2) from their pseudopotential Fermi surface in
1965~\cite{AndersonGold1965}. Our calculation provides its first
principles counterpart for lead, orbit by orbit.

\subsection*{Silicon: absorption, thermal emission and mobility}
\label{sec:semi}

\begin{figure}[t]
\centering
\includegraphics[width=\linewidth]{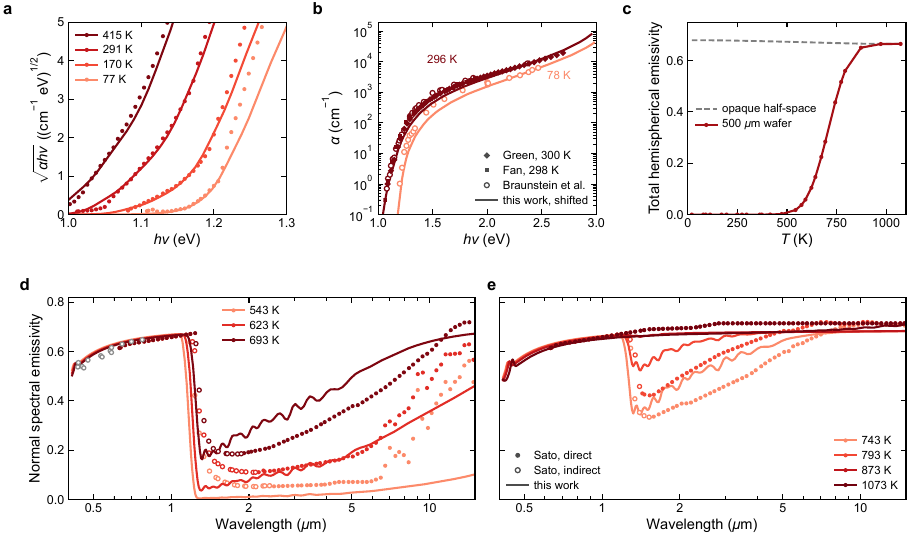}
\caption{\textbf{Absorption and thermal emission of silicon} ($60^3$ $k$, $30^3$ $q$, PBE electrons with the scissor to the Varshni gap). \textbf{a}, Indirect absorption edge at four temperatures, calculated (lines) and measured by Macfarlane \textit{et al.}~\cite{Macfarlane1958} (dots, digitized from the original figure); the model temperatures are 78 and 296~K for the measured 77 and 291~K. \textbf{b}, Absorption coefficient at 78 and 296~K compared with Green~\cite{Green2008} (diamonds, 300~K), Fan~\cite{Fan1956} (squares, 298~K) and Braunstein \textit{et al.}~\cite{Braunstein1958} (open circles); all experimental data are taken from the original publications. In \textbf{a} and \textbf{b} the calculated spectra are rigidly shifted to lower energy by 22--50~meV to match the measured onset at each temperature. \textbf{c}, Calculated total hemispherical emissivity of a 500~$\mu$m wafer and of the opaque half-space. \textbf{d},\textbf{e}, Normal spectral emissivity of the 1.77~mm wafer of Sato~\cite{Sato1967} at the seven measured temperatures: calculated without shift (lines), measured directly (filled) and from transmission and reflection (open); gray symbols refer to visible data at unresolved temperatures.}
\label{fig:si}
\end{figure}

For a semiconductor the same vertex enters the second-order amplitudes
of phonon-assisted absorption~\cite{Noffsinger2012} (Methods). We have
completed the velocity operator with the L\"owdin position dipoles of
the orbital basis, applied the Levine--Allan scissor to the measured
gap with its Varshni temperature dependence, and rescaled the coupling
to the measured dielectric constant with the Resta
model~\cite{Ponce2018}. In Fig.~\ref{fig:si}a,b we compare the
calculated absorption of silicon at the indirect edge and between the
indirect and direct gaps with the measurements of Macfarlane
\textit{et al.}~\cite{Macfarlane1958}, Braunstein \textit{et
al.}~\cite{Braunstein1958}, Fan~\cite{Fan1956} and
Green~\cite{Green2008}. The calculated onset of the indirect edge lies
22--50~meV above the measured one, the difference growing with
temperature (Supplementary Fig.~S2). To facilitate the comparison with
experiment, the calculated spectra of Fig.~\ref{fig:si}a,b have been
rigidly shifted to lower energy by these amounts, in order to match
the onset of the measured absorption at each temperature, as in
Ref.~\cite{Noffsinger2012}. After the shift the absorption coefficient
between 1.1 and 2.5~eV is 0.56--0.73 of the measured one at room
temperature and about one half at 78~K, and at the lowest temperatures
the calculated edge rises more slowly than the measured one
(Fig.~\ref{fig:si}a). These are the known limits of the
independent-particle theory without excitons and multiphonon
processes~\cite{Noffsinger2012,Zacharias2015}. The refractive index is
within 5\% of experiment and the free-carrier absorption of the
intrinsic hot crystal within a factor of 1.0--1.5 (Supplementary
Information).

Assembled, without any shift, into the dielectric function and into
the incoherent slab optics of the 1.77~mm wafer of
Sato~\cite{Sato1967}, these quantities determine its normal spectral
emissivity at the seven measured temperatures between 543 and 1073~K
(Fig.~\ref{fig:si}d,e). The opaque visible window (0.4--1.1~$\mu$m) is
reproduced within $+0.014$ and $-0.005$ at every temperature,
including its slow decrease with $T$, and the whole spectrum from 0.4
to 15~$\mu$m within 0.035 (rms) at 873 and 1073~K, where the free
carriers make the wafer opaque. At 543 and 623~K the calculated wafer
is too transparent between 1.3 and 6~$\mu$m, with a bias of $-0.15$
and $-0.10$. This is due to the edge, which lies 0.07~$\mu$m short,
and to the donors of the specimen: at these temperatures the intrinsic
carrier density is below the donor density, which the intrinsic model
ignores. The multiphonon lattice bands beyond 6~$\mu$m, on the other
hand, are not part of the model.

The same calculation predicts the total hemispherical emissivity at
all temperatures (Fig.~\ref{fig:si}c). For a 500~$\mu$m wafer it stays
below $10^{-6}$ up to 300~K and rises to 0.02 at 543~K, 0.28 at 693~K
and 0.66 at 1073~K, the opaque limit. 

The carrier mobilities of silicon are obtained from the scattering
rates of the same vertex, with the dynamical quadrupoles extracted
from the finite-$\bq$ DFPT densities ($Q_{x,yz} = 14.42\,e$\,bohr$^2$,
compared with 13.66 in Ref.~\cite{Brunin2020}). At 300~K the MRTA
electron mobility is 1300--1350~cm$^2$\,V$^{-1}$\,s$^{-1}$ on
$60^3$--$84^3$ grids without the quadrupoles and 1250--1290 with them,
and the hole mobility 490--540 and 450--500, respectively. The
quadrupole term thus lowers the electron mobility by 4\% and the hole
mobility by 8\% on every grid. These values compare well with the
measured mobilities, $\approx 1400$~cm$^2$\,V$^{-1}$\,s$^{-1}$ for
electrons and 470--500 for holes. The
screening correction of Ref.~\cite{Ponce2018}, which brings $\epsilon$
from 12.97 to 11.94, lowers the mobilities on the quadrupole vertex by
a further 10.0\% for electrons and 14.2\% for holes.

\section*{Discussion}

We have presented a construction of the electron--phonon vertex in the
fixed basis of pseudo-atomic orbitals. In this basis the
variational freedom of Wannier functions is given up in favor of
automation and of a gauge that is defined once, by the projection. The
basis must, however, satisfy a completeness condition: the bands that
enter a property must be well projected. The projectability measures
this condition state by state, and the basis presets (valence shells,
their augmentation by the next shell and the tables generated from the
pseudopotentials) satisfy it for every system studied here, with
projectabilities $p_{n\bk} > 0.95$. Where the condition is satisfied, the
interpolation is as accurate as the Wannier one, as the vertex parity
and the Si absorption demonstrate, and no disentanglement window is
needed for the entangled conduction bands of a metal with spin--orbit
coupling.

The limitations of the present results are those of the underlying
theory rather than of the representation: harmonic phonons at a fixed lattice, the Coulomb pseudopotential $\mu^*$,
the independent-particle optics without excitons or multiphonon
processes. The main advantage of the construction is
its reach: thermal emission, photoemission maps with matrix elements
and orbit-resolved de Haas--van Alphen masses are obtained from the
same interpolated vertex as the gap and the mobility. All these
properties are computed within a single open-source code,
\PAOFLOW{}, which starts from the DFT and DFPT calculations of \QE{}
and is driven by one configuration file per material. The
superconducting, transport, optical and spectroscopic properties of a
material therefore share the same basis and the same vertex, with no
need to pass data between separate codes.
Finally, the same real-space vertex can be obtained from displaced supercells
instead of DFPT, which opens the way to hybrid functionals, DFT+$U(+V)$,
non-periodic reference geometries, thermal cells and alloys within the
same framework.

\section*{Methods}

\subsection*{Electrons in the pseudo-atomic orbital basis}
The Kohn--Sham states $\psi_{n\bk}$ of a DFT calculation are projected
on $N_{\rm awf}$ pseudo-atomic orbitals
$\phi_i$, radial functions times spherical harmonics
generated from the pseudopotential of each element
(\texttt{paoflow-genbasis-ps}). With the projection matrix
$A_{ni}(\bk) = \langle\psi_{n\bk}|\phi_{i\bk}\rangle$, after L\"owdin
orthonormalization of the orbitals, and the projectability
$p_{n\bk} = \sum_i|A_{ni}(\bk)|^2$, the $N_c$ bands with
$p_{n\bk} > p_{\rm thr}$ define the PAO Hamiltonian~\cite{Agapito2013}
\begin{equation}
H(\bk) = A^\dagger_c(\bk)\,\eps_c(\bk)\,A_c(\bk) + \eta\,\big[\,I - A^\dagger_c (A_c A_c^\dagger)^{-1} A_c\big],
\label{eq:Hpao}
\end{equation}
where $A_c(\bk)$ is the $N_c\times N_{\rm awf}$ block of the
projection matrix formed by the rows of these bands, $\eps_c(\bk)$ is
the diagonal matrix of their Kohn--Sham energies, $I$ is the identity
in the orbital space and $\eta$ is an energy that pushes the poorly
projected complement above the energy window. For ultrasoft and PAW pseudopotentials the projections and the
orbital overlaps are taken in the augmented metric
$S = 1 + \sum q_{ij}|\beta_i\rangle\langle\beta_j|$, rotated into the
spinor basis for fully relativistic sets. Once $H(\bR)$ is known from
the Fourier transform on the coarse grid, the Hamiltonian at any $\bk$
can be written as the Wigner--Seitz sum
\begin{equation}
H_{ij}(\bk) = \sum_{\bR} w_{ij}(\bR)\, H_{ij}(\bR)\, \ee^{-\ii\bk\cdot\bR},
\label{eq:Hk}
\end{equation}
with pair-resolved weights $w_{ij}(\bR)$ that place each matrix
element at the image of $\bR + \bm\tau_j - \bm\tau_i$ closest to the
origin. The diagonalization on the dense grid provides $\eps_{n\bk}$
and the eigenvectors $V(\bk)$ in the orbital basis.

\subsection*{The vertex}
The DFPT vertex in the Cartesian displacement basis,
$g^{\kappa\alpha}_{mn}(\bk,\bq) =
\langle\psi_{m,\bk+\bq}|\partial_{\bq\kappa\alpha}V|\psi_{n\bk}\rangle$,
is evaluated natively from the DFPT outputs. It is the sum of
the induced Hartree and exchange--correlation potential
(contracted from the displacement patterns to
Cartesian displacements), the bare local pseudopotential term
$-\ii(\bq+\mathbf G)\cdot\mathbf u_\kappa V^{\rm
loc}_\kappa(|\bq+\mathbf G|)\ee^{-\ii(\bq+\mathbf
G)\cdot\bm\tau_\kappa}$ and the Kleinman--Bylander derivative terms of
the nonlocal pseudopotential, with the spinor coupling coefficients of
the fully relativistic projectors. The irreducible $q$ points of the
DFPT run are unfolded to their stars by rotating the scalar potentials
on the FFT grid, and a DFT calculation on the irreducible wedge is unfolded
by rotating its Bloch functions, with SU(2) and time reversal for
spinors. Both procedures reproduce the explicit full-grid
results to $10^{-7}$--$10^{-12}$. Additionally, the coarse coupling computed by
\EPW{} can be read as an alternative source and shares the gauge of
the projections. The vertex is rotated to the orbital basis with
Eq.~\eqref{eq:gpao} and transformed to
\begin{equation}
g^{\rm PAO}_{ij}(\bR_e,\bR_p) = \frac{1}{N_k N_q}\sum_{\bk\bq} \ee^{-\ii\bk\cdot\bR_e}\ee^{-\ii\bq\cdot\bR_p}\, g^{{\rm PAO}}_{ij}(\bk,\bq),
\label{eq:gR}
\end{equation}
where $N_k$ and $N_q$ are the numbers of points
of the coarse $\bk$ and $\bq$ grids of the DFT and DFPT calculations. The real-space vertex is then
interpolated with Wigner--Seitz weights on both lattices,
\begin{equation}
g^{\rm PAO}_{ij}(\bk,\bq) = \sum_{\bR_e\bR_p} w_{ij}(\bR_e)\,w_{ij}(\bR_p)\,\ee^{\ii\bk\cdot\bR_e}\ee^{\ii\bq\cdot\bR_p}\,g^{\rm PAO}_{ij}(\bR_e,\bR_p),
\label{eq:ginterp}
\end{equation}
and rotated to the dense-grid bands and modes,
\begin{equation}
g_{mn\nu}(\bk,\bq) = \sum_{ij}\sum_{\kappa\alpha} V^*_{im}(\bk+\bq)\; g^{{\rm PAO},\kappa\alpha}_{ij}(\bk,\bq)\;\frac{e^{\nu}_{\kappa\alpha}(\bq)}{\sqrt{M_\kappa}}\; V_{jn}(\bk),
\label{eq:gband}
\end{equation}
where $e^{\nu}_{\kappa\alpha}(\bq)/\sqrt{M_\kappa}$ are the
mass-scaled eigenvectors of the dynamical matrix interpolated from the
coarse force constants with the acoustic sum rule.

The Fourier phase of the dense Hamiltonian follows the
convention of \PAOFLOW{},
$H(\bk) = \sum_\bR \ee^{-2\pi\ii\bk\cdot\bR}H(\bR)$, so that electrons
and vertex are evaluated at the same $\bk$. In practice, the dense $q$
grid is reduced to its irreducible wedge with star weights. The pair
coupling of the anisotropic equations and the Fermi-surface averages
sum the irreducible $q$ points with their multiplicities, which is
exact because $g_{S\bk,S\bk+S\bq} = g_{\bk,\bk+\bq}$.

\subsection*{Finite differences in the fixed basis}
The finite-difference vertex replaces the DFPT potentials with
displaced-supercell calculations. The PAO Hamiltonian of an
$n\times n\times n$ supercell is rebuilt for the reference cell and
for the cells with the reference atom displaced by $\pm\delta$ along
each Cartesian axis, with the orbitals centered at the reference
positions in every case. In this way $\partial_{\kappa\alpha}V_{ij} =
[H^{\rm PAO}_{ij}(+\delta) - H^{\rm PAO}_{ij}(-\delta)]/2\delta$ is
the derivative of the Kohn--Sham potential in the fixed orbitals,
exact within the span of the $N_c$ well-projected bands up to the
$O(\delta^2)$ error. A basis that followed the atoms would add the
basis-rotation commutator $[H, \partial U]$, which is not part of the
vertex. The completion term of Eq.~\eqref{eq:Hpao} depends on the
displacement through the projector
$P = A_c^\dagger(A_cA_c^\dagger)^{-1}A_c$ on the well-projected bands,
and its contribution, $-\eta\,\partial P$, is removed by projecting the
derivative with the projector of the reference cell, $P_0\,
\partial_{\kappa\alpha}V\,P_0$. The supercell derivative
is unfolded to the primitive grids, $\bk$ on the supercell $k$ points
and their $n^3$ images and $\bq$ on the $n^3$ grid of the cell, and
Fourier-transformed to $g(\bR_e,\bR_p)$ in the layout of the DFPT
vertex, from where the interpolation and every driver are the same.
For large cells, random all-atom displacements (Gaussian patterns of
every atom and their mirrors, with a locality cutoff and a ridge
regularization of the per-element fit) replace the Cartesian set, and
the forces of the same runs provide the force constants and the
phonons of the cell.

\subsection*{Long-range fields and screening}
The long-range dipole (Fr\"ohlich) and quadrupole
vertex~\cite{Verdi2015,Sjakste2015,Brunin2020}
\begin{equation}
g^{\mathcal L}_{\kappa\alpha}(\bq) = \ii\frac{4\pi e^2}{\Om}\sum_{\mathbf G\ne-\bq}\frac{\ee^{-(\bq+\mathbf G)^2/4\alpha}}{(\bq+\mathbf G)\cdot\eps^\infty\cdot(\bq+\mathbf G)}\,\ee^{-\ii(\bq+\mathbf G)\cdot\bm\tau_\kappa}\Big[(\bq+\mathbf G)\cdot Z^*_\kappa - \tfrac{\ii}{2}(\bq+\mathbf G)\cdot Q_\kappa\cdot(\bq+\mathbf G)\Big]_\alpha
\label{eq:lr}
\end{equation}
is diagonal in the orbitals in the PAO gauge; in silicon the Born
charges vanish and only the quadrupole term contributes. It is
subtracted at every coarse $(\bk,\bq)$ before the transform to
$(\bR_e,\bR_p)$ and restored at every dense $\bq$. The dipole--dipole
and quadrupole terms of the dynamical matrix are treated in the same
way. In particular, the dynamical quadrupoles are fitted to
the finite-$\bq$ DFPT densities of twelve small
$\bq$~\cite{Brunin2020}. Following Ponc\'e \textit{et
al.}~\cite{Ponce2018}, the DFT overscreening of the perturbation in a
semiconductor is corrected by rescaling every matrix element by the
ratio of the Resta model dielectric functions~\cite{Resta1977} with
the DFT and the measured $\eps(0)$.

\subsection*{Eliashberg functions and the Migdal--Eliashberg equations}
With $N_k$ and $N_q$ the numbers of points of the dense $\bk$ and
$\bq$ grids, $w_q$ the weights of the irreducible dense $q$ grid, $N(E_F)$ the
density of states per spin and the Fermi-surface double delta
evaluated with Gaussian widths $\sigma$ or with the adaptive widths
$a|\nabla_\bk\eps|\Delta k$ of Yates \textit{et al.}~\cite{Yates2007},
the mode-resolved coupling and the Eliashberg function can be written
as
\begin{gather}
\lambda_{\bq\nu} = \frac{1}{N(E_F)\,\omega_{\bq\nu}^2}\,\frac{1}{N_k}\sum_{\bk mn}|g_{mn\nu}(\bk,\bq)|^2\,\delta(\eps_{n\bk}-E_F)\,\delta(\eps_{m,\bk+\bq}-E_F),\\
\alpha^2F(\omega) = \tfrac12\sum_{\bq\nu}w_q\lambda_{\bq\nu}\omega_{\bq\nu}\delta(\omega-\omega_{\bq\nu}),
\end{gather}
while $\lambda_{\rm tr}$ and $\alpha^2F_{\rm tr}$ carry the weight
$\mathbf v_{n\bk}\cdot(\mathbf v_{n\bk}-\mathbf v_{m\bk+\bq})/\langle
v^2\rangle_{FS}$~\cite{Allen1971}. $T_c$ follows from the McMillan and
Allen--Dynes formulas~\cite{McMillan1968,AllenDynes1975}. The
isotropic Migdal--Eliashberg equations are solved on the Matsubara
axis with Anderson mixing and continued to the real axis by Pad\'e
approximants~\cite{Vidberg1977} and by the
Marsiglio--Schossmann--Carbotte equations~\cite{Marsiglio1988}. $T_c$
is taken both from the closure of the gap and from the temperature at
which the largest eigenvalue of the linearized kernel crosses one, as
in \EPW{}~\cite{Margine2013}. The anisotropic equations act on the
pair coupling $\Lambda_{ab}(\omega)$ of the irreducible Fermi-surface
states within $\pm0.2$~eV of $E_F$, which is accumulated during the
loop over irreducible $q$ points, with Coulomb weights restricted to
the pairs that the phonons couple and a separable (SVD) Matsubara
kernel. As a check, for a single state they reduce to the isotropic
equations to $10^{-14}$.

\subsection*{Transport, optics and thermal emission}
For metals the resistivity is obtained from Allen's variational
formula with $\alpha^2F_{\rm tr}$ and the Fermi-surface plasma
frequency~\cite{Allen1971,Grimvall1981}. The frequency-dependent
conductivity is the extended Drude form with Allen's optical
scattering rate and mass enhancement~\cite{Allen1971,Holstein1964}, to
which the interband dielectric function of the extended basis may be
added. Fresnel's equations then provide the reflectivity and the
optical constants and, via Kirchhoff's law, the spectral directional
emissivity, whose Planck-weighted hemispherical average is the total
emissivity. The state-resolved rates
\begin{equation}
\begin{split}
\frac{1}{\tau_{n\bk}} = {}&\frac{2\pi}{\hbar N_q}\sum_{\bq m\nu}\frac{|g_{mn\nu}(\bk,\bq)|^2}{2\omega_{\bq\nu}}\Big[(n_{\bq\nu}+f_{m\bk+\bq})\delta(\eps_{m\bk+\bq}-\eps_{n\bk}-\omega_{\bq\nu})\\
&+ (n_{\bq\nu}+1-f_{m\bk+\bq})\delta(\eps_{m\bk+\bq}-\eps_{n\bk}+\omega_{\bq\nu})\Big]
\end{split}
\label{eq:serta}
\end{equation}
(SERTA; MRTA weights each process by $1-\mathbf v_{n\bk}\cdot\mathbf
v_{m\bk+\bq}/|\mathbf v_{n\bk}|^2$) enter the relaxation-time
conductivity, the drift mobilities at the chemical potential of charge
neutrality and the Seebeck
coefficient~\cite{Giustino2017,Ponce2018,Claes2022}. For
semiconductors, on the other hand, the direct dielectric function and
the phonon-assisted one, with the two interfering second-order
amplitudes of Noffsinger \textit{et al.}~\cite{Noffsinger2012} and the
occupation factors that include stimulated emission, are assembled
into $\eps(\omega,T)$ with the Kramers--Kronig real part, the
Levine--Allan velocities of the scissor~\cite{LevineAllan1989} and the
Varshni gap~\cite{Varshni1967}. The velocity operator is $\mathbf v =
\partial_\bk H + \ii[H,\bm\tau] + \ii[H,\mathbf d]$, where the
L\"owdin position dipoles $\mathbf d(\bk)$ of the orbital basis bring
the interband $|v|^2$ of Si within 10\% of the exact plane-wave
values. The spectral emissivity of a free-standing wafer follows from
Kirchhoff's law and the incoherent sum of internal
reflections~\cite{Modest2013}.

\subsection*{Self-energies and photoemission}
The Engelsberg--Schrieffer self-energy $\Sigma(\omega,T)$ of
$\alpha^2F$~\cite{Engelsberg1963}, its state-resolved version on the
pair coupling of the Fermi-surface states, and the Fan--Migdal
self-energy $\Sigma_{n\bk}(\omega,T)$ on a dense $q$ grid dress the
spectral function $A_{n\bk}(\omega)$. The photoemission intensity is
$\sum_n|M_{n\bk}|^2A_{n\bk}(\omega)f(\omega)$ convolved with the
resolution, where $M_{n\bk}$ are the dipole matrix elements of the PAO
orbitals to a plane-wave final state in the formalism of
\textsc{chinook}~\cite{Day2019}, which we have implemented for the
real-harmonic and spinor PAO bases. $\lambda$ is read from the slope
of $\mathrm{Re}\,\Sigma$ at $E_F$.

\subsection*{Computational details}
All calculations have been performed with \QE{} and \PAOFLOW{} within
the generalized gradient approximation (PBE)~\cite{PBE1996}. The
input files of \QE{} have been generated with \texttt{paoflow-gen-qe},
the command-line tool of the \PAOFLOW{} distribution that writes them
from an entry of the AFLOW database
(\url{https://aflow.org})~\cite{2012-Curtarolo-CMS-58-227}, from which
the crystal structures are taken (Si AUID: aflow:08ab41c5f54850db; Pb
AUID: aflow:bea2f327a8d40908). For lead
we have used PseudoDojo norm-conserving
pseudopotentials~\cite{Hamann2013,vanSetten2018}, scalar relativistic
(v0.5) and fully relativistic (v0.4), a cutoff of 56~Ry
and a $22^3$ k-point grid with a Methfessel--Paxton smearing of
0.04~Ry~\cite{Methfessel1989}. The DFPT has been carried out on $6^3$
$q$ grids, and the coupling has been built natively from the
DFPT potentials on a $12^3$  grid (20 and 40 bands)
with the $6^3$ $q$ grid. The PAO basis includes the 5$d$, 6$s$, 6$p$,
6$d$, 7$s$ and 7$p$ shells (18 orbitals, 36 spinors) with a
projectability threshold of 0.95. The dense grids range from
$48^3/24^3$ to $72^3/36^3$ and $60^3/60^3$, with a smearing of 0.05~eV
(0.03 and 0.08~eV checked), $\mu^* = 0.10$ and a Matsubara cutoff of
0.1~eV.

For silicon we have used a cutoff of 38~Ry with PseudoDojo norm-conserving
pseudopotentials, scalar relativistic
(v0.5), $12^3/6^3$ coarse grids
with 21 bands, a 26-orbital basis (44 orbitals for the direct
dielectric function), $60^3/30^3$ dense grids for the optics and
$60^3$--$84^3$ for the mobilities.

\subsection*{Use of AI tools}
Parts of the code development, debugging and documentation, the
scripts that generate the figures and the editing of the manuscript
have been carried out with the assistance of the large language model
Claude Opus 5.5 (Anthropic), used through GitHub Copilot and Claude
Code in Visual Studio Code. The scientific choices, i.e., the methods,
the physical approximations, the computational parameters and the
interpretation of the results, have been made by the author, who has
verified the code, the data and the text and takes full responsibility
for the content of the paper.

\section*{Data availability}
The inputs and outputs of the calculations of this paper can be recreated using PAOFLOW's own \texttt{paoflow-gen-qe} and \texttt{paoflow-gen}, or from the integrated browser-based GUI, \texttt{PAOFLOW-Studio}, on the AFLOW entries and following the electron-phonon system workflow. The data and scripts that generate every figure in this paper are archived at https://doi.org/10.5281/zenodo.24515135.

\section*{Code availability}
\PAOFLOW{} is open source (GPL-3) and available at
\url{https://github.com/marcobn/PAOFLOW}. The electron--phonon
workflows, the system folders of Pb and Si with their drivers, and the
generator \texttt{paoflow-gen elphon} are also in the
\texttt{examples/elphon} directory of the repository.

\section*{Acknowledgements}
MBN wishes to thank Anooja Jayaraj, Marcio Costa, Jagoda S{\l}awi{\'n}ska, and Davide Ceresoli for enlightening and stimulating discussions. This work has been supported, in part, by the National Science Foundation under grant No. DMR2523217 and by an Anthropic AI for Science grant. Computational resources have been provided by the Texas Advanced Computing Center of the University of Texas, Austin.

\section*{Author contributions}
M.B.N. designed the study, implemented the methods, performed the
calculations and wrote the manuscript.

\section*{Competing interests}
The author declares no competing interests.

\bibliographystyle{unsrtnat}
\bibliography{references}
\end{document}